\documentclass[a4paper,11pt]{article}
\usepackage{style/pos}
\usepackage[separate-uncertainty=true, range-units=single]{siunitx}
\usepackage{xspace}
\usepackage{amssymb}

\newcommand{\eV}[1]{$10^{#1}\,$eV\xspace}

\newcommand{\Xmax}{\ensuremath{X_\mathrm{max}}\xspace}
\newcommand{\Xrit}{\ensuremath{X_\mathrm{RIT}}\xspace}

\newcommand{\gcm}{\ensuremath{\,\mathrm{g}\,\mathrm{cm}^{-2}}\xspace}
\newcommand{\Fig}[1]{Fig.\ \ref{fig:#1}\xspace}

\title{
Expected performance of interferometric air-shower measurements with radio antennas}
\ShortTitle{Interferometric air-shower measurements}

\author*[a,b]{Felix Schlüter}
\author[a,c]{Tim Huege}

\affiliation[a]{Karlsruhe Institute of Technology, Institute for Astroparticle Physics, Karlsruhe, Germany}

\affiliation[b]{Universidad Nacional de San Martín, Instituto de Tecnologías en Detección y Astropartículas,\\ Buenos Aires, Argentina}

\affiliation[c]{Vrije Universiteit Brussel, Astrophysical Institute, Brussels, Belgium}

\emailAdd{felix.schlueter@kit.edu, tim.huege@kit.edu}

\abstract{
Interferometric measurements of the radio emission of extensive air showers allow reconstructing cosmic-ray properties. A recent simulation study with an idealised detector promised measurements of the depth of the shower maximum \Xmax with an accuracy better than \SI{10}{g\,cm^{-2}} \cite{rit}. In this contribution, we evaluate the potential of interferometric \Xmax measurements of (simulated) inclined air showers with realistically dimensioned, sparse antenna arrays. We account for imperfect time synchronisation between individual antennas and study its inter-dependency with the antenna density in detail. 
We find a strong correlation between the antenna multiplicity (per event) and the maximum acceptable inaccuracy in the time synchronisation of individual antennas. From this result, prerequisites for the design of antenna arrays for the application of interferometric measurements can be concluded. 
For data recorded with a time synchronisation accurate to \SI{1}{ns} within the commonly used frequency band of \SIrange{30}{80}{MHz}, an antenna multiplicity of $\gtrsim 50$ is needed to achieve an \Xmax reconstruction with an accuracy of \SI{20}{g\,cm^{-2}}. 
This multiplicity is achieved measuring inclined air showers with zenith angles $\theta \geq 77.5^\circ$ with \SI{1}{km} spaced antenna arrays, while vertical air showers with zenith angles $\theta \leq 40^\circ$ require an antenna spacing below \SI{100}{m}. Furthermore, we find no improvement in \Xmax resolution applying the interferometric reconstruction to measurements at higher frequencies, i.e., up to several hundred MHz.
}

\FullConference{37$^{\rm{th}}$ International Cosmic Ray Conference (ICRC 2021)\\
		July 12th -- 23rd, 2021\\
		Online -- Berlin, Germany}

\begin{document}
\maketitle

\section{Introduction}
The radio signal from extensive air showers is the superposition of electromagnetic radiation emitted by mostly the showers' electrons and positrons. Interferometric techniques exploit the coherence in signals received by multiple observers and can be used to reconstruct the properties of air showers. While standard in radio astronomy, such technique have only been applied with limited success for the reconstruction of air shower properties, such as the depth of shower maximum \Xmax, with sparse arrays of antennas with a wide field-of-view \cite{Heino, Jandt, Apel:2021oco}. Recently, the so-called radio-interferometric technique (RIT) was proposed in reference \cite{rit}, predicting reconstruction of \Xmax with a resolution of \SI{3}{g\,cm^{-2}} (\SI{10}{g\,cm^{-2}}) for inclined (vertical) air showers. 


Here, we perform a simulation study to investigate whether these promising results can be transferred to the reconstruction of inclined air showers using a more realistic detector, i.e., using realistically dimensioned, sparse antenna-arrays with imperfect time synchronisation between the antennas. The time synchronisation is key to an accurate interferometric reconstruction. We thus study in particular the inter-dependency between the maximum tolerable time jitter, i.e., inaccuracy in the time synchronisation, and the antenna multiplicity. This work mainly refers to the radio emission from 30$\,$MHz to 80$\,$MHz which is commonly used by most current radio air-shower experiments and in particular by the Radio Detector of the Pierre Auger Observatory which will consists of 1661 radio antennas on an area of 3000$\,$km$^{2}$ \cite{PontIcrc2019}. However we also evaluate the technique for higher frequencies ranges which will be accessible by radio air-shower experiments such as GRAND. Our study has been published in \cite{rit_schlueter}.

\section{Air shower simulations with various detector layouts}
To study the expected RIT performance under realistic conditions, we use two sets of CoREAS v7.7401 \cite{HuegeCoREAS2012} simulations of inclined air showers. The simulations are performed, without loss of generality, for the ambient conditions of the Pierre Auger Observatory. This entails, among other things, the atmospheric model (density and refractivity profile) and the observation altitude of \SI{1400}{m} a.s.l.. The particle cascades of the extensive air showers were simulated with QGSJetII-04, UrQMD, and an optimized thinning level of $10^{-6}$.

The first simulation set contains 50 proton showers with antennas situated on a {\it dense}, flat hexagonal grid with an antenna spacing of \SI{250}{m}. The showers have a zenith angle of \SI{77.5}{^\circ}, two different azimuths angles \SI{0}{^\circ} (east) and \SI{30}{^\circ} (east-north), respectively, and an energy of \eV{18.4}. The dense grid of simulated antennas allows to define various sub-arrays of antennas with different spacings. Thus the reconstruction can be tested for different detector layouts with varying numbers of antennas included in the reconstruction ($\equiv$ antenna multiplicity). For a given antenna spacing, different sub-arrays with different centers of gravity, corresponding to different impact positions of the shower in the array, can be defined. This is summarized in Tab.\ \ref{tab:sim}
\begin{table}
  \caption{Different detector layouts, i.e., antenna spacings, evaluated here. Number of different sub-arrays for the specific antenna spacing $n_\mathrm{rec}$ for all 50 showers combined and the average antenna multiplicity $\langle n_\mathrm{ant} \rangle$ for these sub-arrays.}
  \centering\vspace{0.2cm}
  \begin{tabular}{c|cccccc}
    \\[-1em]
    spacing / m & 250 & 500 & 750 & 1000 & 1250 & 1500 \\\hline
    $n_\mathrm{rec}$ & 50 & 200 & 450 & 800 & 1250 & 1800 \\
    $\langle n_\mathrm{ant} \rangle$ & 1342 & 336 & 149 & 84 & 54 & 37
  \end{tabular}
  \label{tab:sim}
\end{table}

The second simulation set has 1902 proton- and iron-induced showers simulated with a \SI{1.5}{km} hexagonal antenna array and isotropically distributed arrival directions for zenith angles from 65$^\circ$ to 85$^\circ$. The shower energies range from \eV{18.4} to \eV{20.1}, uniformly randomized in $\log_{10}(E / \text{eV})$. The shower impact point at ground, in the following called ``core'', is randomly distributed within a finite \SI{3000}{km^2} array. For each shower all antennas are simulated for a maximum geometry-dependent distance to the shower axis (cf.\ Tab.\ \ref{tab:sim2} for the maximum antenna-axis distance and average antenna multiplicity).
\begin{table}
  \caption{Maximum antenna-axis distance (measured perpendicular to the shower axis, i.e., in the shower plane) and average (simulated) antenna-multiplicity for the 1.5 km hexagonal grid as a function of the zenith angle in \SI{2.5}{\degree}-bins.}
  \centering\vspace{0.2cm}
  \begin{tabular}{c|cccccccc}
    \\[-1em]
$\theta / ^\circ$ & 66.25 & 68.75 & 71.25 & 73.75 & 76.25 & 78.75 & 81.25 & 83.75 \\\hline
$\langle n_\mathrm{ant} \rangle \pm \sigma_\mathrm{ant} $ & 9 $\pm$ 1 & 10 $\pm$ 1 & 11 $\pm$ 1 & 16 $\pm$ 3 & 27 $\pm$ 6 & 47 $\pm$ 11 & 87 $\pm$ 21 & 173 $\pm$ 42 \\
$r_\mathrm{ant}^\mathrm{max}$ / m & 1500 & 1500 & 1508 & 1822 & 2230 & 2785 & 3563 & 4707
  \end{tabular}
  \label{tab:sim2}
\end{table}
\section{Interferometric measurement of the shower maximum}
\begin{figure}[tbp]
    \centering
    \includegraphics[width=\linewidth]{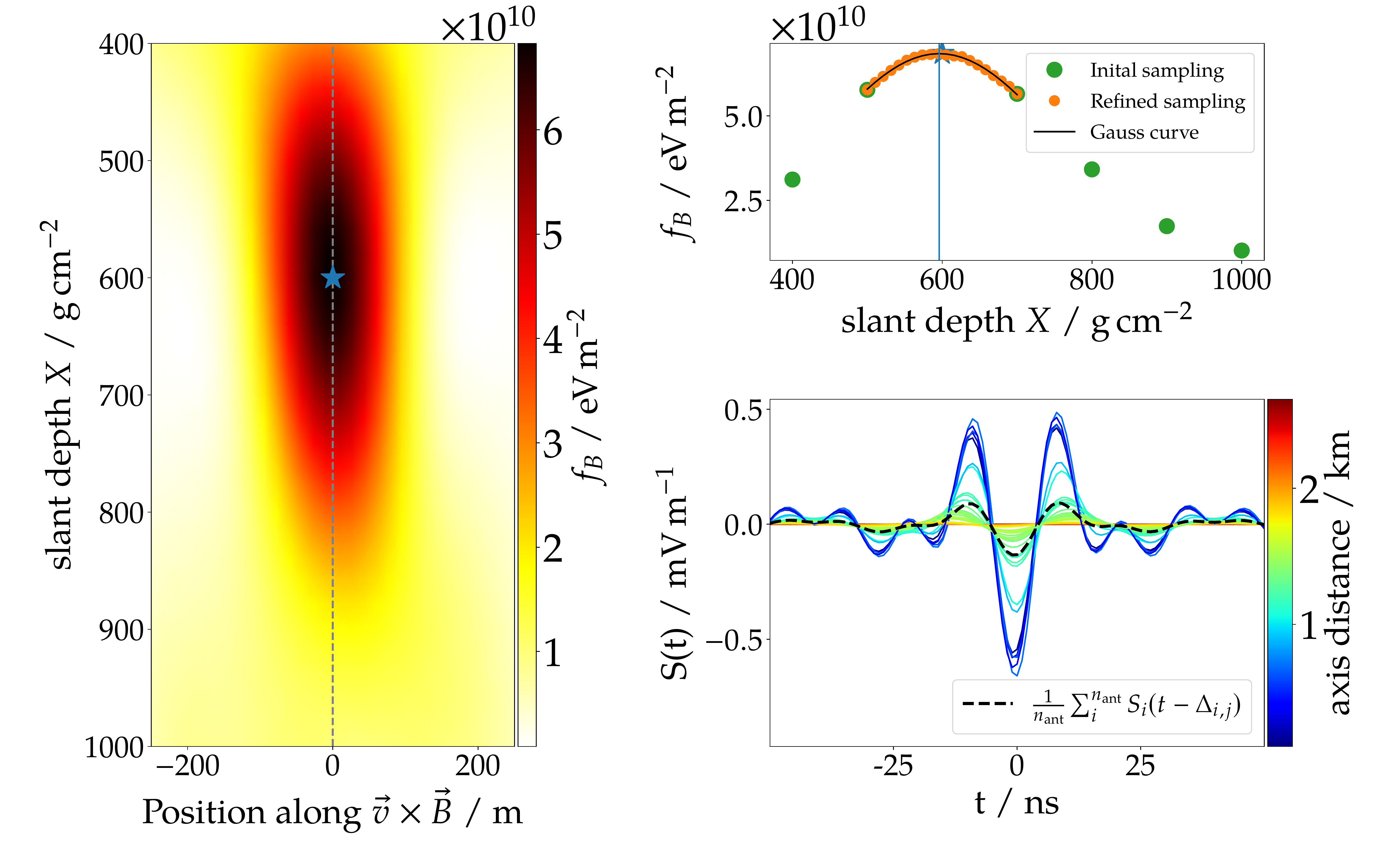}
    \caption{
    Cross section of the longitudinal (y-axis) and lateral (x-axis) profile of the coherent energy fluence $f_{B_j}$ (color coded) of a \SI{2.51}{EeV}, \SI{77.5}{\degree} proton shower sampled with 37 antennas on a \SI{1500}{m} grid along the shower axis (vertical dashed line). Longitudinal profile $f_{B_{j}}(X)$ along the MC shower axis for the same shower (right top). Time shifted signals at ground $S_i(t - \Delta_{i,j})$ for a location $\vec{j}$ at \SI{600}{g\,cm^{-2}} on the shower axis (blue star).
    }
    \label{fig:tomo}
\end{figure}
The reconstruction algorithm which is shorty summarized here, is explained in detail in \cite{rit_schlueter} and is based on the one in reference \cite{rit}. The prime idea is to determine the beam-formed signal $B_j$ which is received by several observers at ground $\vec{i}$ from an arbitrary position $\vec{j}$ in the atmosphere. Thereby 3d-profiles of the beam-formed signal, ``interferometric maps'', are generated, which can be interpreted as depicting the origin of the \emph{coherent} radio emission released during the development of the particle cascade. From these interferometric maps (cf. \Fig{tomo} ({\it left}), the shower axis and the shower maximum \Xmax are reconstructed. 

$B_j$ is calculated via a sum over all time-shifted signals measured at ground $S_i$
\begin{equation}
    \label{eq:rit}
    B_{j}(t) = \sum_i^{n_\mathrm{ant}} S_{i}(t - \Delta_{i,j}), \: \: \text{with} \: \Delta_{i,j} = \frac{d_{i,j}\overline{n_{i,j}}}{c_0}.
\end{equation}
The time shift $\Delta_{i,j}$ corresponds to the light-propagation time along the distance $d_{i,j}$ of a spherically expanding wave through the refractive atmosphere with an average refractive index $\overline{n_{i,j}}$. In \Fig{tomo} ({\it bottom right}) the time-shifted signals $S_i(t - \Delta_{i,j})$ are shown for a position $\vec{j}$ on the shower axis of an example shower. The calculation is performed along straight lines\footnote{Also in CoREAS the propagation is computed along straight trajectories, which is found to accurately describe the coherence between radio emission released in different parts of the shower development \cite{Schlueter_2021}}. The refractive index profile $n(h)$ follows the Gladstone-Dale law with an $n(0) = 1 + 3.12 \cdot 10 ^ {-4}$ as also adopted in the CoREAS simulation \cite{rit_schlueter}. To sample the interferometric maps, we integrate over $B_j^2(t)$ in a \SI{100}{ns} window around the largest signal peak and obtain the energy deposit per unit area, i.e., the coherent energy fluence $f_{B_j} / \,$eV$\,$m$^{-2}$:
\begin{equation}
    \label{eq:rit3}
    f_{B_{j}} = \epsilon_0 \, c \, \Delta t \sum_{t_\mathrm{peak}-50\mathrm{ns}}^{t_\mathrm{peak}+50\mathrm{ns}} B^2_j(t)
\end{equation}
where $\epsilon_0$ is the vacuum permittivity and $c$ the speed of light in vacuum. 

For the signals measured at ground $S_i(t)$ we only use the radio emission in the $\vec{v} \times \vec{B}$-polarisation\footnote{Determined using the MC arrival direction.} and between 30 to 80$\,$MHz. For inclined air showers the emission in the $\vec{v} \times \vec{B}$-polarisation is mostly comprised of the dominant geomagnetic emission. We found no apparent correlation between the shower development and the $f_{B_j}$ profile for the emission in the $\vec{v} \times (\vec{v} \times \vec{B})$-polarisation which is only comprised of the sub-dominant charge-excess emission. 

\begin{figure}[tbp]
    \centering
    \includegraphics[width=.60\linewidth]{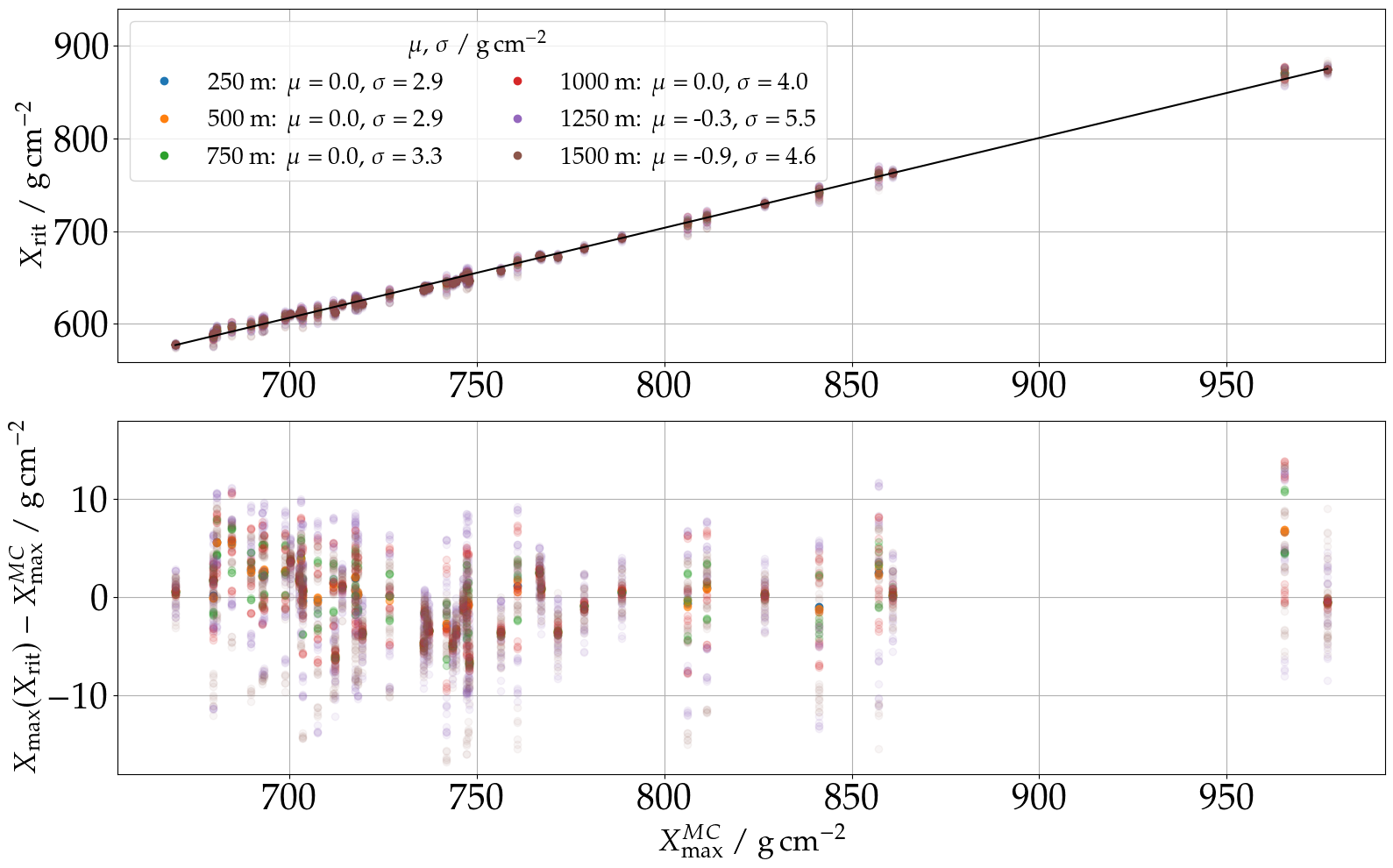}\hfill
    \includegraphics[width=.37\linewidth]{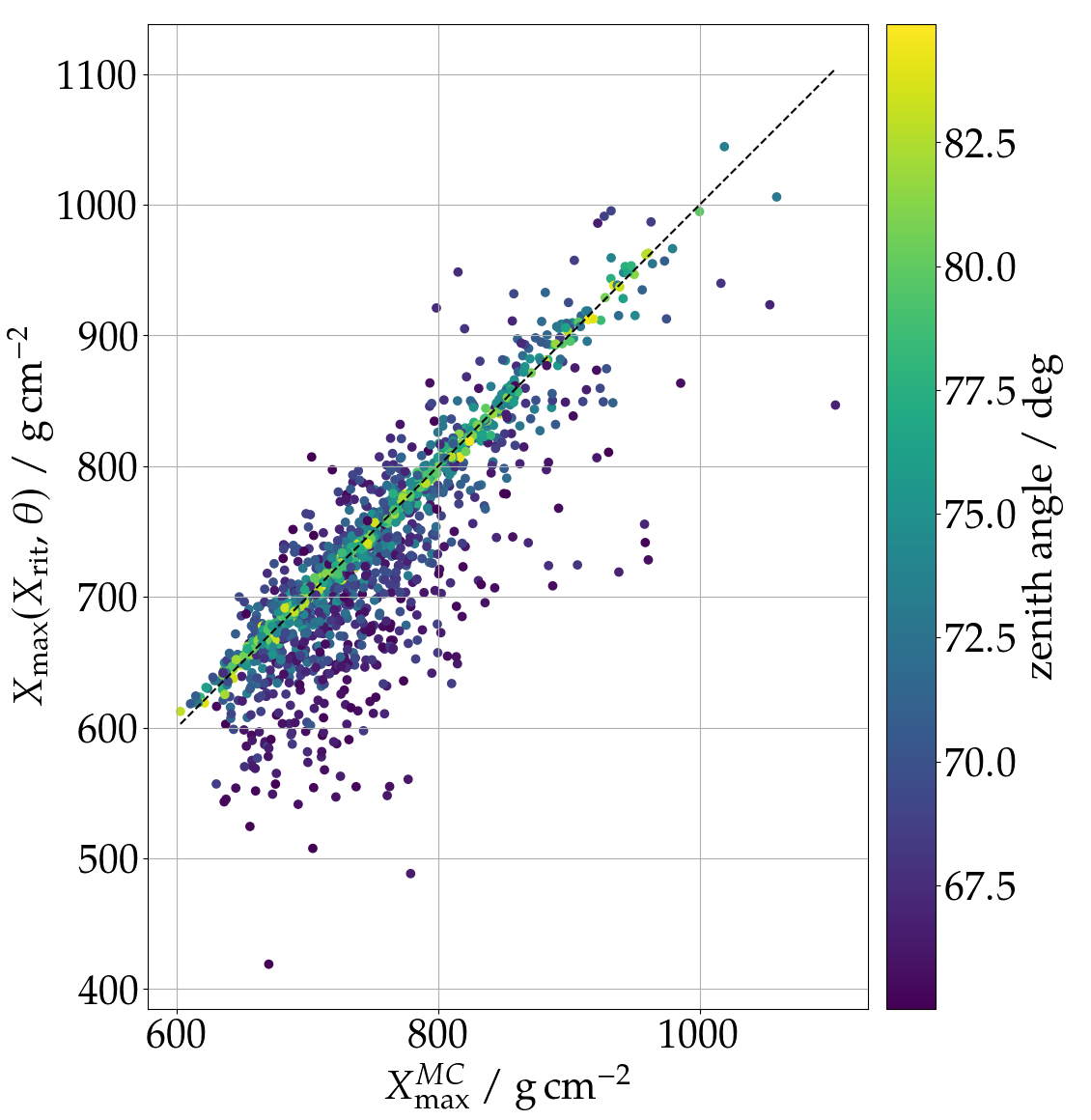}
    \caption{Reconstruction with perfect time synchronisation between the antennas and along the MC shower axis. {\it Left}: Correlation between $X_\mathrm{RIT}$ and $X_\mathrm{max}^\mathrm{MC}$ for the dense simulations for different (sub-)arrays with different antenna-spacings (color-coded). The black line indicates the calibration curve according to Eq.\ \eqref{eq:xmax}. Residuals are shown between reconstructed and true depth of the shower maximum \Xmax. The legend illustrates the reconstruction bias and resolution for different array spacings. {\it Right}: Correlation between reconstructed \Xmax and $X_\mathrm{max}^\mathrm{MC}$ for the simulations on the 1.5 km grid (according to Eq.\ \eqref{eq:xmax2}). The black dashed line indicates identity, the color code denotes the zenith angle.}
    \label{fig:aera1}
\end{figure}

In \Fig{tomo} ({\it left}) cross section along the shower axis, sampled in $f_{B_j}$, of an example shower is shown. It is apparent that the maximum of the $f_{B_j}$ profile lies on the shower axis (dashed line at $\vec{v} \times \vec{B} = 0$). The longitudinal profile of $f_{B_j}$ along the true shower axis is shown in the same figure ({\it top right}). It exhibits a clear maximum which is defined as \Xrit. This maximum is found by fitting a Gauss curve to the sampled profile. For a given zenith angle, a linear relationship between \Xrit and \Xmax is found which allows reconstructing the latter. The correlation between \Xrit and \Xmax for showers from the {\it dense} simulation set, at a zenith angle of 77.5$^\circ$, is shown in \Fig{aera1} ({\it left}) and described by 
\begin{equation}
\label{eq:xmax}
    X_\mathrm{max}(X_\mathrm{RIT}) = 1.03 \cdot X_\mathrm{RIT} + 76.15\,\text{g}\,\text{cm}^{-2}.
\end{equation}
The residual illustrates a very accurate reconstruction with $\sigma_{\Xmax} \lesssim 5\,$g$\,$cm$^{-2}$ and without a significant dependency on the antenna spacing / multiplicity respectively. 

\Fig{aera1} ({\it right}) shows the interferometric reconstruction of \Xmax for the showers of the second simulation set with isotropic arrival directions in the zenith angle region between 65$^\circ$ and 85$^\circ$. The color code indicates the zenith angle and demonstrates that only for zenith angles $\gtrsim 72.5^\circ$ an accurate reconstruction is possible. This is owed to the fact that for showers with a smaller zenith angle, the antenna multiplicity on the 1.5 km antenna grid, which on average is below 12, is too low (cf. Tab.\ \ref{tab:sim} and \Fig{auger2} ({\it right})). To extend Eq.\ \eqref{eq:xmax} to the reconstruction of showers with zenith angles other than 77.5$^\circ$ it is sufficient to add a linear zenith-angle dependence to the offset between \Xrit and \Xmax
\begin{equation}
    \label{eq:xmax2}
    X_\mathrm{max}(X_\mathrm{RIT}, \theta) =  1.04 \cdot X_\mathrm{RIT} + \left(68.31 - \frac{\theta - 77.5^\circ}{0.35^\circ} \right)\text{g}\,\text{cm}^{-2}.
\end{equation}
To derive the parameters of Eq.\ \eqref{eq:xmax2}, only showers with a zenith angle greater than 75$^\circ$ were used. As we will see later that, above a certain antenna multiplicity ($\sim$ 20, cf.\ \Fig{auger2} ({\it right}, $\sigma_t = 0\,$ns)), the accuracy is rather independent of the zenith angle (antenna multiplicity).

For the results shown in \Fig{aera1}, the longitudinal $f_{B_j}$-profiles were sampled along the MC shower axis and the beam-formed signals were calculated with perfect time synchronisations between the antennas. This allows to establish a calibration for \Xrit (Eqs.\ \eqref{eq:xmax} and \eqref{eq:xmax2}), which we found to be not dependent on the detector layout, i.e., antenna spacing The here achieved results are comparable and thus confirm the results from reference \cite{rit}. In the following section, the \Xmax reconstruction is executed and evaluated under more realistic conditions, i.e., with imperfect time synchronisation between the antennas and along an imperfect, reconstructed shower axis.

The shower axis can be reconstructed with RIT in a similar fashion as \Xmax. The procedure is described in \cite{rit_schlueter}. With perfect time synchronisation, the shower axis can be reconstructed with very high accuracy with a resolution on the opening angle w.r.t MC shower axis of $\sigma_{68\%} < 0.01^\circ$ in most cases. It seems that, depending on the azimuth angle, the accuracy remains relatively constant until the station density drops below a certain value.

\section{Accuracy of the interferometric reconstruction of the shower maximum using a realistic detector}
\begin{figure}[tbp]
    \centering
    \includegraphics[width=.49\linewidth]{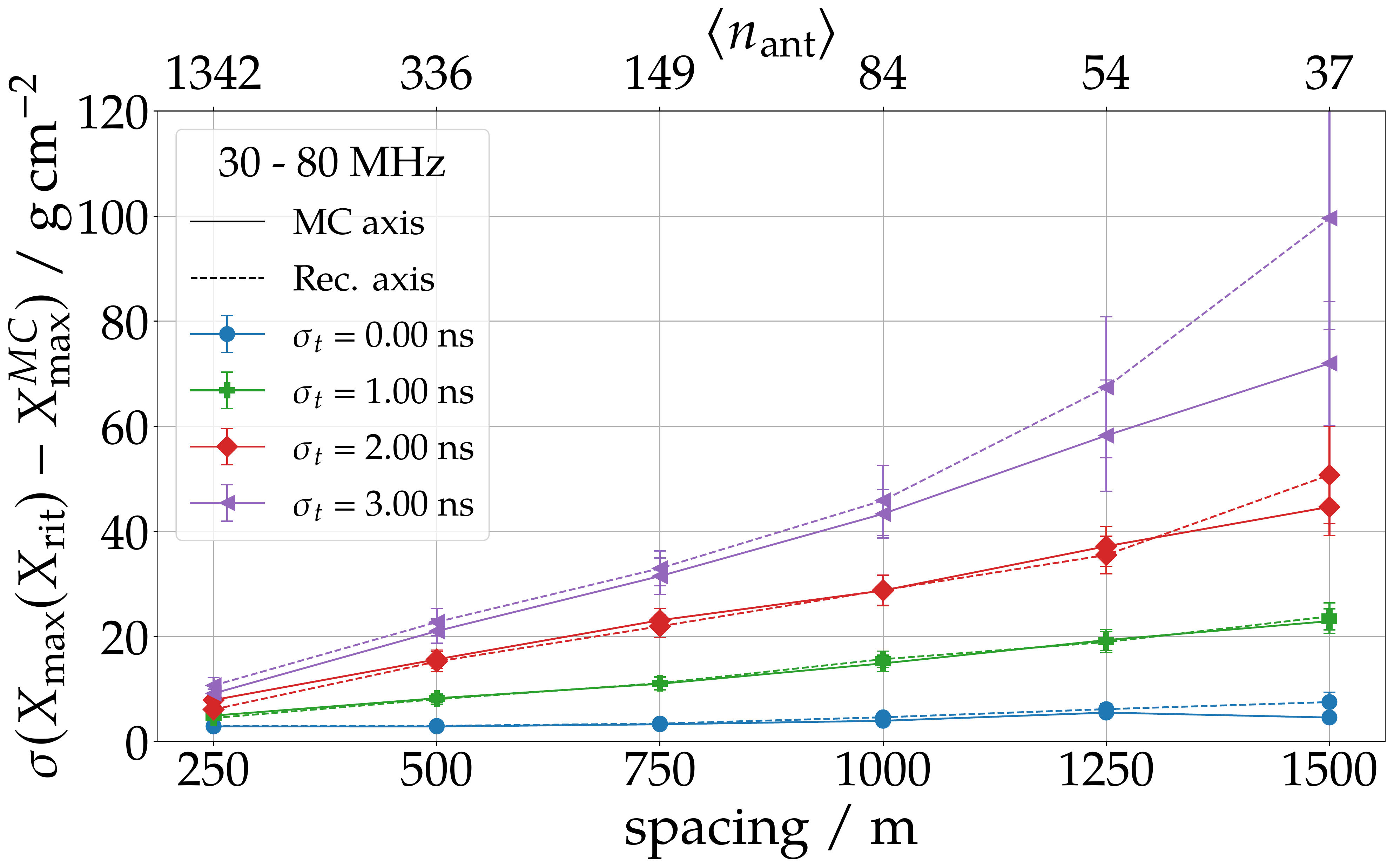}\hfill
    \includegraphics[width=.49\linewidth]{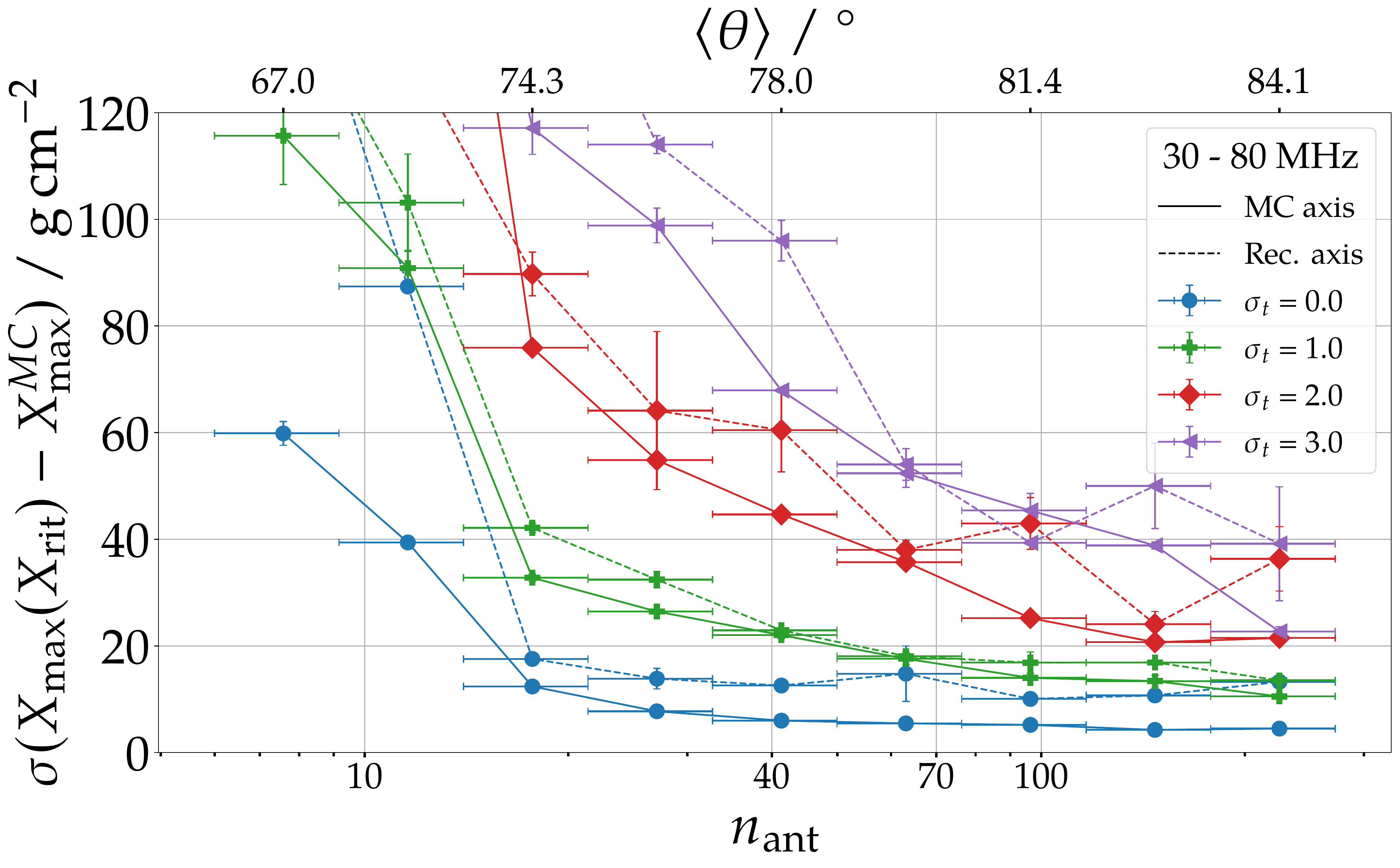}
    \caption{{\it Left}: Reconstruction resolution in \Xmax of the 50 densely sampled showers with a zenith angle of \SI{77.5}{\degree}. Resolution is shown for different time jitter scenarios (different colors \& markers) and along the MC shower axis (solid line) or reconstructed axis (dashed line) as a function of the antenna spacing. {\it Right}: \Xmax-Resolution binned as a function of the antenna multiplicity for the simulations on the 1.5 km grid along the true and reconstructed shower axes (solid and dashed lines, respectively), and for different time jitters. The horizontal error bars indicate the bin size, the vertical bars correspond to the statistical fluctuation of the resolution. A few outliers in the axis reconstruction causing the visible fluctuations in the \Xmax resolution.}
    \label{fig:auger2}
\end{figure}
Now we perform the reconstruction assuming an imperfect time synchronisation between the antennas. For this purpose we randomly add a normally-distributed time jitter $\mathcal{N}(0, \sigma_t)$ to the timing of each antenna. Furthermore, we perform the reconstruction twice, along the MC and along a RIT-reconstructed shower axis. In \Fig{auger2} ({\it left}) the \Xmax resolution for the dense simulations is shown as a function of the antenna spacing for the different sub-arrays (x-axis) and different time jitter scenarios (color coded). The average antenna multiplicity for showers measured with a certain antenna spacing is shown on the top x-axis. It is clearly visible that the negative impact of the imperfect time synchronisation increases with increasing antenna spacing / decreasing antenna multiplicity. While for the showers measured with the very dense \SI{250}{m}-grid ($\langle n_\mathrm{ant} \rangle = 1342$), the resolution only mildly deteriorates with the magnitude of the time jitter and remains better than \SI{10}{g\,cm^{-2}}, the resolution degrades significantly for showers measured on a \SI{1500}{m}-grid ($\langle n_\mathrm{ant} \rangle = 37$) resulting in a impractical resolution of \SI{100}{g\,cm^{-2}} for a time jitter of $\sigma_t = 3\,$ns. It is also shown that the reconstruction along the reconstructed shower axis is not degrading the \Xmax reconstruction in most cases. The reconstruction of the showers with varying zenith angles on a 1.5\,km grid yields a similar picture, cf.\ \Fig{auger2} ({\it right}). The impact of a time jitter on the reconstruction clearly depends on the antenna multiplicity. For this simulation set with a finite antenna array, the discrepancy in \Xmax resolution between the reconstruction along the simulated and reconstructed shower axis can be larger than for simulations from the set of showers which were always well centered in the middle of the array. This is caused by the fact that the footprints of the showers which hit the ground close to the edge of the array are not evenly sampled. This disturbs the axis reconstruction and consequently the \Xmax reconstruction.

With imperfect time synchronisation we see an overall decrease in resolution, for a 3$\,$ns timing resolution, the resolution decreases by a factor of 2 - 3 with respect to perfect timing \cite{rit_schlueter}.

So far the analysis of the RIT reconstructions used the radio signal in the 30 to 80$\,$MHz band. To test whether it is beneficial to include higher frequencies we evaluate the RIT reconstruction also for the two frequency bands 50 to 200$\,$MHz and 150 to 350$\,$MHz. In \Fig{freqs1} ({\it left}) the reconstruction for perfect time synchronisation is shown for the 3 mentioned frequency bands, which shows no advantage in the reconstruction of \Xmax at higher frequencies. The accuracy seems to degenerate for the reconstruction of \Xmax along the reconstructed shower axis and for larger antennas spacings at higher frequencies. We find that the distribution of $f_{B_j}$ around the shower axis in the interferometric maps (cf.\ \Fig{tomo} ({\it left})) becomes narrower for higher frequencies and that grating lobes, i.e., local maxima, become more prominent. With the algorithm for the reconstruction of the shower axis described in reference \cite{rit_schlueter} and employed here, we found no advantage applying RIT to higher frequencies. In addition higher frequencies demand more stringent coherence criteria. That means that the maximum tolerable time jitter is lower for higher frequencies \cite{rit_schlueter}. 

\section{Generalization to lower zenith angle and other considerations}
\begin{figure}[tbp]
    \centering
    \includegraphics[width=.51\linewidth]{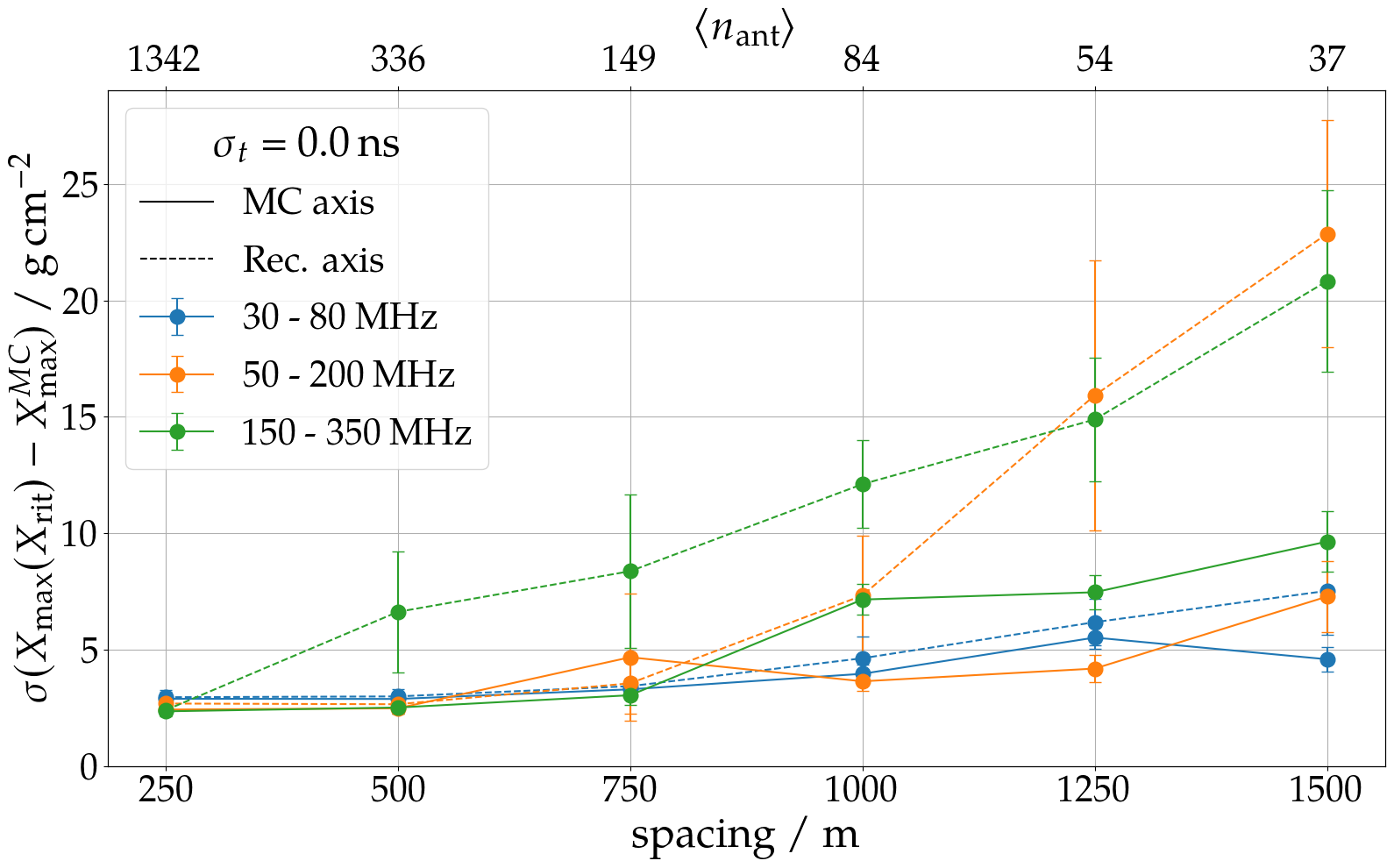}\hfill
    \includegraphics[width=.47\linewidth]{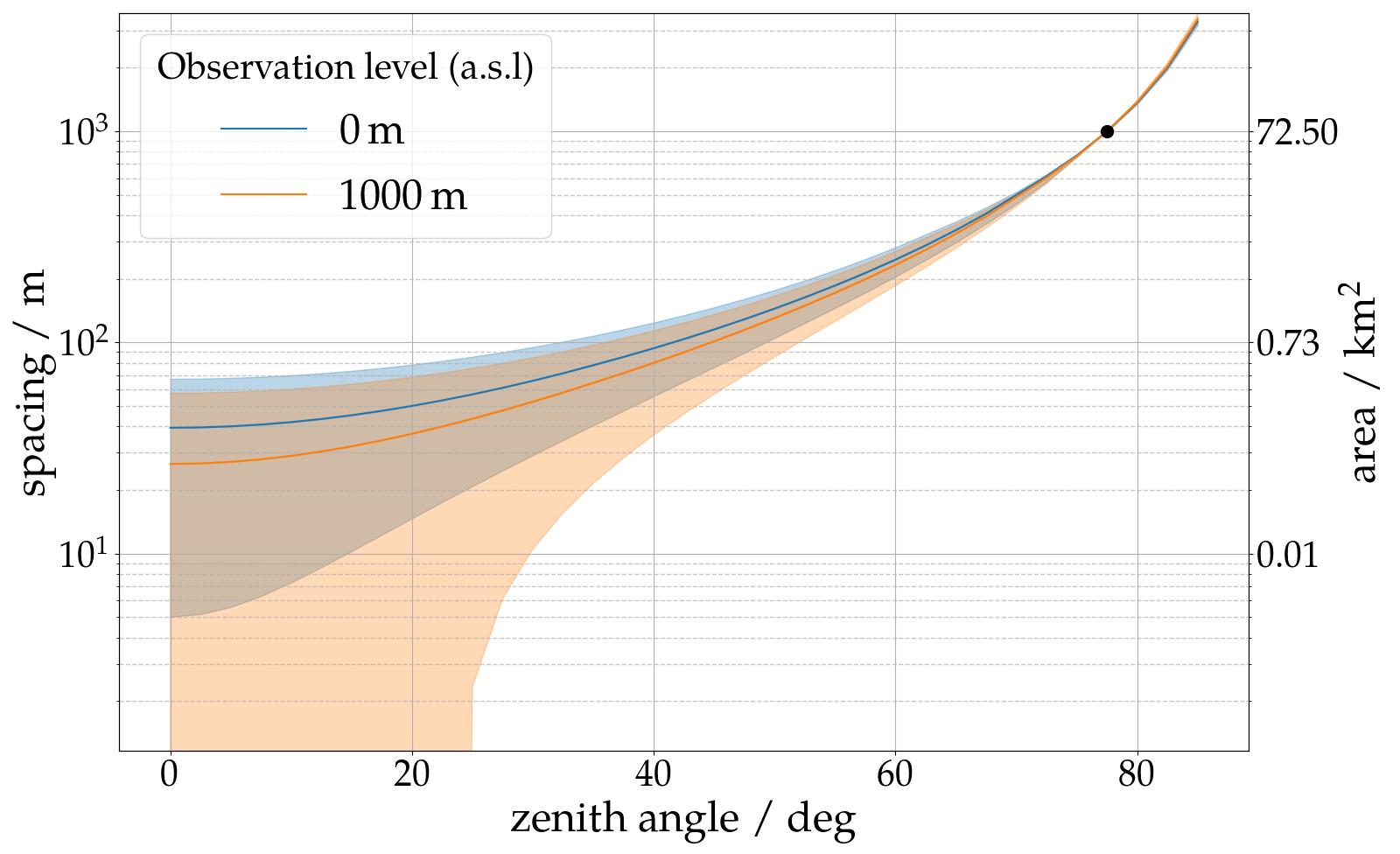}
    \caption{{\it Left}: \Xmax-resolution for the frequency bands \SIrange{30}{80}{MHz},  \SIrange{50}{200}{MHz}, and \SIrange{150}{350}{MHz} (different colors) along the MC or reconstructed shower axis (solid or dashed line) for perfect time synchronisation as a function of the antenna spacing. {\it Right}: Antenna spacing as function of the zenith angle required to achieve an antenna multiplicity of $\sim \,$84. This is the mean antenna multiplicity for showers with $\theta$ = \SI{77.5}{\degree} measured with a \SI{1000}{m} hexagonal grid (black dot). The resolution achieved with these reference showers and a time jitter of $\sigma_t = 1\,$ns is  $\sigma_{\Xmax} = 16\gcm$}.
    \label{fig:freqs1}
\end{figure}

\label{sec:diss}
The study presented here is limited to inclined air showers and, in case of arrays with an antenna spacing of < \SI{1.5}{km}, to a zenith angle of $\theta = 77.5^\circ$. To extrapolate our results to less inclined air showers we assume that the \Xmax resolution solely depends on the antenna multiplicity. Hence we model the size of the radio-emission footprint $A \sim \pi r_\mathrm{che}^2 / \cos \theta$ to estimate the antenna multiplicity as a function of the zenith angle (assuming that it will depend purely on shower geometry but not energy). $r_\mathrm{che}$ is the radius of the Cherenkov cone (defined in the shower frame). For a reference point, given by the previous analysis and chosen here as $\sigma_{\Xmax}(\theta = 77.5^\circ, \Delta_\mathrm{ant} = 1000\,\mathrm{m}, \sigma_t = 1\,\mathrm{ns}) = 16\gcm$, corresponding to $\langle n_\mathrm{ant} \rangle = 84$, we can estimate how the antenna spacing needs to change as a function of the zenith angle to keep the same accuracy (antenna multiplicity). This is shown in \Fig{freqs1} ({\it right}) for two different observation levels (color-coded) and \Xmax values ranging from \SIrange{550}{950}{g\,cm^{-2}} with a mean value of \SI{750}{g\,cm^{-2}} (solid lines). To keep the accuracy of a \SI{1000}{m}-spaced antenna for showers with a zenith angle of 77.5$^\circ$, the antenna spacing needs to decrease below \SI{100}{m} for vertical showers with $\theta < 40^\circ$.

In reality, the antenna multiplicity will also be affected by the triggering and data acquisition system of the considered experiment (while the procedure in this study resembles more a trigger-less readout).

\section{Conclusions}
\label{sec:conc}
We studied the reconstruction accuracy of the depth of the shower maximum \Xmax with the radio-interferometric-technique RIT for realistically dimensioned air-shower detector arrays considering an imperfect time synchronisation between the antennas. We found a clear correlation between the maximal tolerable inaccuracy in the time synchronisation allowing an accurate \Xmax reconstruction and the number of antennas participating in the reconstruction. For the 30--80\,MHz band, an accurate reconstruction of \Xmax ($\sigma_{\Xmax} \lesssim 20\gcm$) is possible with a time synchronisation of \SI{1}{ns} or better and a sufficiently large number of antennas per shower ($\gtrsim 100$). Such a level of accuracy in the time synchronisation is challenging to achieve for sparse arrays with wirelessly communicating antenna arrays. Thus it seems unlikely that these criteria can be met by existing or currently planned experiments such as the Pierre Auger Observatory or GRAND which are designed without special considerations for interferometry. However, experiments with a large number of antennas and very accurate time synchronisation such as the Square Kilometer Array have great potential to exploit interferometric measurements of \Xmax. 

We found no improvement in accuracy applying the interferometric reconstruction to data recorded with higher frequencies while at the same time a more accurate time synchronisation between antennas is needed. This stresses the importance for future experiment to also encompass the recording of low frequent signals down to the tens of megahertz.\\

\noindent
\begin{minipage}[t]{0.58\textwidth}

\end{minipage}\hfill
\begin{minipage}[t]{0.4\textwidth}
\acknowledgments
\vspace{-7pt}
\footnotesize
We are thankful to M.\ Gottowik for the thoughtful reading of our manuscript. Felix Schlüter is supported by the Helmholtz International Research School for Astroparticle Physics and Enabling Technologies (HIRSAP) (HIRS-0009). Simulations for this work were performed on the supercomputer BwUniCluster 2.0 and ForHLR II at KIT funded by the Ministry of Science, Research and the Arts Baden-Württemberg and the Federal Ministry of Education and Research.
\end{minipage}

\end{document}